\documentclass[10pt,letterpaper,twocolumn]{article} %% two column, final layout

\usepackage{ol2}
\usepackage[draft,implicit=false]{hyperref}
\usepackage{amsmath}

\begin{document}

\twocolumn[ %% activate for two-column option

% Titre
\hyphenation{Sensor}

\title{Variation around the Pyramid theme: optical recombination and optimal~use of photons}

%% For REVTeX it is possible to automate superscript and e-mail callouts with the superscriptaddress option; see REVTeX4 documentation.

\author{Olivier Fauvarque,$^{1,*}$ Benoit Neichel,$^1$ Thierry Fusco$^{1,2}$ and Jean-Francois Sauvage$^{1,2}$}

\address{$^1$Aix Marseille Universit\'e, CNRS, LAM (Laboratoire d'Astrophysique de Marseille) UMR 7326, 13388, Marseille, France
\\
$^2$ONERA--the French Aerospace Laboratory, F-92322 Ch$\hat{a}$tillon, France \\

$^*$Corresponding author: olivier.fauvarque@lam.fr
}

% Résumé

\begin{abstract} We propose a new type of Wave Front Sensor (WFS) derived from the Pyramid WFS (PWFS). This new WFS, called the Flattened Pyramid-WFS (FPWFS), has a reduced Pyramid angle in order to optically overlap the four pupil images into an unique intensity. This map is then used to derive the phase information. In this letter this new WFS is compared to three existing WFSs, namely the PWFS, the Modulated PWFS (MPWFS) and the Zernike WFS (ZWFS) following tests about sensitivity, linearity range and low photon flux behavior. The FPWFS turns out to be more linear than a modulated pyramid for the high-spatial order aberrations but it provides an improved sensitivity compared to the non-modulated pyramid. The noise propagation may even be as low as the ZWFS for some given radial orders. 
Furthermore, the pixel arrangement being more efficient than for the PWFS, the FPWFS seems particularly well suited for high-contrast applications.
\end{abstract}

% REPLACE WITH CORRECT OCIS CODES FOR YOUR ARTICLE
                          % NOTE: \ocis{} IS ALIASED TO \pacs{} BUT MUST
                          % FORMAT THE TERMS CORRECTLY FOR EACH JOURNAL

 ] %% activate for two-column option

%\section{Introduction}
The aim of a Wave Front Sensor (WFS) is to code the phase information using an incoming photon flux. WFSs are usually divided into two main classes, those working in the pupil plane like e.g. the Shack-Hartmann \cite{Sh71}, and those performing this operation by imaging the pupil after a Fourier filtering in the focal plane.
The Foucault's knife (see \cite{Wil75}) is a first example of this latter approach. In that case, the focal plane mask is simply an Heaviside function, with half of the light passing through, while the other half is blocked. By re-imaging a pupil plane, one can derive the phase information by analyzing the intensity distribution.
This principle has later been generalized by Ragazzonni et al. \cite{Rag96} with the \emph{Pyramid}-WFS (PWFS). In the case of the PWFS the filtering is made by a transparent square pyramid, its summit being positioned on the focal point of the optical system.  The prism shape of this optical object spreads the complex amplitude into four parts coding differently the spatial frequencies. The differential information computed from the four resulting intensities allows to easily code the phase information.

\begin{figure}[htb]
\centerline{\includegraphics[width=3.5cm]{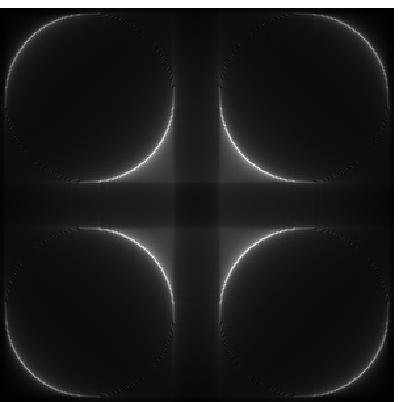}~~\includegraphics[width=3.5cm]{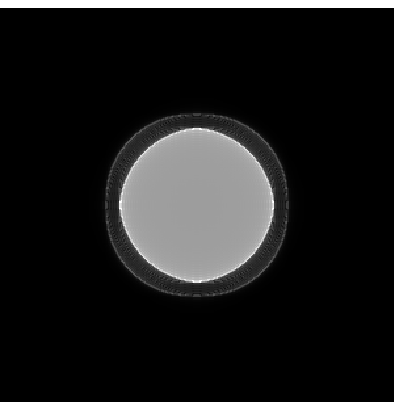}}
\caption{Intensity on the detector for a circular pupil with no aberration for two different Pyramid angles. The overlap rate of the four pupil images equals to: 0\% for the left image which is the usual PWFS case and 90\% for the right one which is the typical FPWFS case. \label{rappr}}
\end{figure}

We propose here a variation around the Pyramid concept, the idea being to recombine the phase information before the detection. To do so, we propose to reduce significantly the pyramid angle so that the four images, which were, in the original concept, completely separated, move closer and overlap. This is illustrated in Fig. \ref{rappr} where we show two configurations: a full pupil separation (left insert; classical PWFS configuration) and an overlap rate of 90\% (right insert). This resulting intensity distribution is then used as a phase information coding.
Due to the shape of the Pyramid, this new WFS will be called FPWFS for Flattened Pyramid Wave Front~Sensor.

The FPWFS considered here will use a constant overlap rate of 90\%. This choice results from a first trade-off between the main performance criteria introduced in this letter, namely sensitivity and linearity range. Note that we found that the influence of this parameter regarding these performance criteria was not critical, and will be detailed in a next paper.\\

The aim of this letter is to derive the main characteristics of this new WFS in terms of linearity, sensitivity and noise propagation, and to compare them with existing WFSs. To do so, three WFSs have been chosen, respectively: 
\begin{itemize}
\item the PWFS already introduced above.
\item the Modulated PWFS (introduced by Ragazzoni in the same paper \cite{Rag96}) which improves the PWFS linearity range, as a trade-off with sensitivity. 
In this letter, a circular tip/tilt modulation with an amplitude going from 1 to 6 $\lambda/D$ (in radius) will be considered.
\item the Zernike WFS (ZWFS), introduced by N'Diaye in \cite{Ndia13}, and known as a reference sensor in term of noise propagation as illustrated by Guyon in \cite{Guy05}.
\end{itemize}

In order to study the performance of a WFS, the first step is to know how to compute the intensity directly coming from the detector to create a quantity (called meta-intensity) which should be linear with the Optical Path Difference (OPD).
We assume that the incoming flux is uniform (i.e. we neglect scintillation effects) and defined by $\psi = \sqrt{n} P e^{2\imath\pi \frac{\delta}{\lambda} }$ where $\psi$ is the complex amplitude of the light just in the pupil, $n$ is the number of incoming photons in the full pupil, P is the indicator function of the pupil which informs about the geometry of the pupil and $\delta$ is the OPD of the incoming light. The aim of any WFS is to measure this latter quantity.
The intensity on the detector $I(\delta,n)$ depends on the OPD and is proportional to the flux.\\

\begin{figure}[htb]
\centerline{\includegraphics[width=3.5cm]{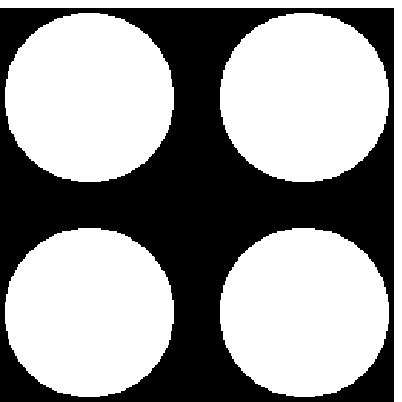}~~\includegraphics[width=3.5cm]{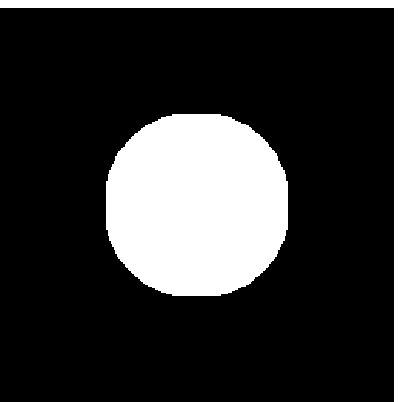}}
\caption{Valid pixels on the detector for the PWFS and MPWFS (left) and FPWPS (right). Note the non-circular shape of the Flattened Pyramid's mask.\label{mask}}
\end{figure}

In the case of the PWFS and the MPWFS, the meta-intensities usually defined (see \cite{Rag96}) are the classical \emph{slopes} maps $S_x$ and $S_y$ defined by:
\begin{align}
S_x=\frac{I_1+I_2-I_3-I_4}{I_1+I_2+I_3+I_4}~~{\rm and}~~S_y=\frac{I_3+I_2-I_1-I_4}{I_1+I_2+I_3+I_4}
\end{align}
where $I_1$, $I_2$, $I_3$ and $I_4$ are the intensity in the 4 pupil images. Note that this meta-intensities could be chosen differently, see \cite{Ver04}. Only the photons inside the geometrical pupil footprint are taken into account. This is illustrated by the mask shown in the left image of Fig. \ref{mask}. 

Concerning, the FPWFS and the ZWFS, the chosen meta-intensity consists in a return-to-reference and a normalization:
\begin{align}
mI(\delta) &= \frac{I(\delta,n)-I(ref,n)}{I(ref,n)}
\end{align}
$I(ref,n)$ is the intensity on the detector when the optical aberration corresponds to the reference phase. 
Such a meta-intensity, although basic, is the simplest way to recode the intensity and can be used for every WFS which optically codes the phase into intensity. 
In our case, the reference phase is the zero-OPD but in practice, it may contain WFS path aberrations. 
In addition, and in order to follow the same computation as for the PWFS,
photons which are diffracted outside the geometrical footprint of the pupil on the detector are removed from computation. For the FPWFS, it corresponds to the pupils overlap area, illustrated by the right image of Fig. \ref{mask}. 
For the ZWFS, this area is simply the unique geometrical pupil image (see N'Diaye et al. \cite{Ndia13}).\\

In this first \emph{performance test} paragraph, we will focus on the \textbf{sensitivity} of the FPWFS in terms of noise propagation. Following the results of Rigaut and Gendron \cite{Rigaut92}, the diagonal elements of $(M^ tM)^{-1}$, where $M$ is the interaction matrix of the system, have to be studied.
The interaction matrix is defined on the first 24 Zernike radial orders which corresponds to the first 299 Zernike modes. Each mode has the same amplitude: 10 nm RMS at $\lambda$=630 nm (which corresponds to a 0.1 rad RMS phase amplitude). Moreover, we will use a very high number of pixels on the detector (i.e. 4096) so that any effects of pupil sampling on the noise propagation coefficients, i.e. aliasing effects, can be neglected. 
%Note that this is only a numerical constraint linked to the simulation, not a hardware constraint. 
For each WFS case, the interaction matrix is built with the meta intensities defined above ($S_x$ and $S_y$ for the PWFS and the MPWFS; mI for the FPWFS and the ZWFS).

\begin{figure}[htb]
\centerline{\includegraphics[width=8cm]{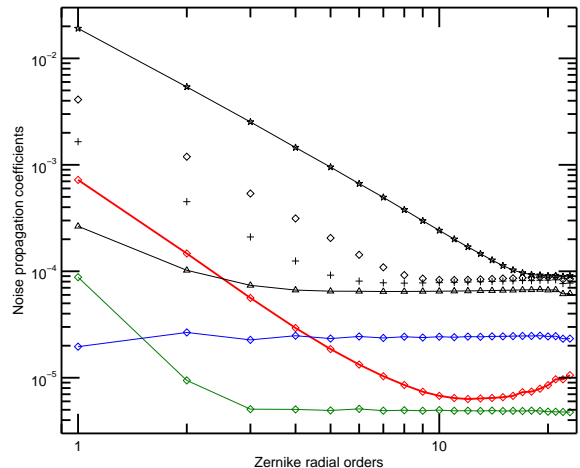}}
\caption{Noise propagation of the FPWFS (red), ZWFS (green), PWFS (blue) and MPWFS (black: $\star_{6\lambda/D}$, $\diamond_{3}$, $+_{2}$ and  $\triangle_{1}$) . \label{noi}}
\end{figure}

Fig. \ref{noi} shows the diagonal elements of $(M^tM)^{-1}$ averaged for each Zernike radial order. 
In terms of noise propagation, the ZWFS proves to be the most  optimal WFS as shown by Guyon \cite{Guy05} and N'Diaye \cite{Ndia13}.
The PWFS presents a very good performance, although five times larger than the previous one, with a flat distribution of the errors across the Zernike modes. 
The modulation increases the value of the noise propagation coefficients for higher frequencies up to five times the PWFS one.
Modulation also changes the slope for lower frequencies. 
This is in good agreement with the findings of Verinaud et al. in \cite{Ver04} and Guyon in \cite{Guy05}. 
In comparison, the FPWFS has a very interesting behavior, with a noise propagation close to the MPWFS (modulation of 1.5 $\lambda$/D) for low order Zernikes (up to a radial order of $\sim$4), and then an improved performance when compared to the PWFS itself. 
The noise propagation coefficients are even almost as low as those of the ZWFS for Zernike radial orders in the range of 10 to 20.
Finally, the noise propagation performance of the FPWFS is at least 10 times better (which corresponds to a magnitude gain of 2.5) than the MPWFS modulated at 6 $\lambda/D$ which is the typical operating modulation amplitude.

% \\

\begin{figure}[htb]
\centerline{\includegraphics[width=8cm]{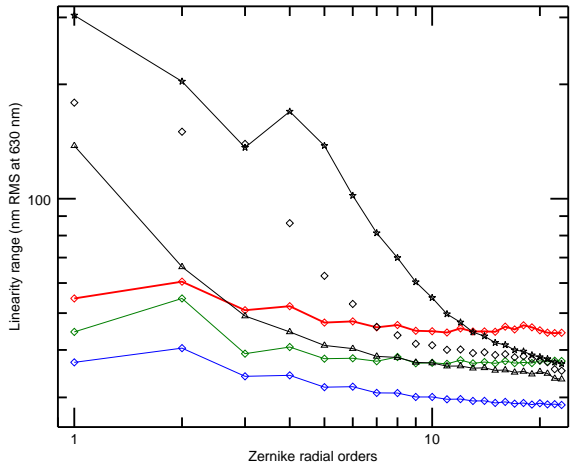}}
\caption{Linearity range of the FPWFS (red), ZWFS (green), PWFS (blue) and MPWFS (black: $\star_{6 \lambda/D}$, $\diamond_{3}$ and $\triangle_{1}$).\label{dyn}}
\end{figure}

The following paragraph is focused on the \textbf{linearity range} of the FPWFS when compared to the other WFSs. The numerical simulations performed here test the validity of the interaction matrix of each WFS when the aberration amplitude is not equal to the calibration amplitude (here equal to 10 nm RMS at $\lambda$=630 nm). 
To do this, for each Zernike mode, an aberration with an increasing amplitude is sent to the WFSs. The outgoing meta-intensity is inverted via the pseudo-inverse of the interaction matrix. 
The linearity range associated to this mode is defined as the amplitude where the error between the estimated and the injected amplitude reaches 5 nm RMS. This is shown in Fig. \ref{dyn} for the 24 first Zernike radial orders.

As already demonstrated in previous studies (Burvall et al. in \cite{Bur06}), the PWFS has a low level of linearity range for all considered modes. This level stays quite flat and slowly decreases with the radial order. 
The FPWFS and the ZWFS have the same type of flat linearity range. Nevertheless, the FPWFS is better than the two others since its mean linearity range stays aroung 50 nm RMS compared to 40 nm for the ZWFS and 30 nm for the PWFS.

Concerning the MPWFS, the tip/tilt modulation releases, as expected, the linearity limitation of the PWFS, especially for the low frequencies. Unfortunately, this gain dramatically drops for the high radial orders. As a consequence the FPWFS becomes better from a certain radial order: for a modulation of 1$\lambda/D$, this order equals to 3, for 3$\lambda/D$ it equals to 7 and for a modulation of 6$\lambda/D$ the radial order of equivalent linearity range between the FPWFS and the MPWFS equals to 17.

In the context of Extreme Adaptive Optics, this result could imply that the FPWFS may not need modulation device.

The previous study was only interested in the linearity range regarding isolated aberrations. But the WFSs have to analyse real turbulent phase screens which correspond to a sum of many aberrations. In order to know how each WFS behaves regarding this coupling of several aberrations modes, the \emph{bootstrap} test consists in trying to close an adaptive optics loop when a typical turbulence screen is injected in the Adaptive Optics (AO) system.
First simulations show that the MPWFS does manage to close the loop and is the fastest WFS, whereas the ZWFS usually diverges. Despite its low linearity range, the PWFS turns out to boostrap effictively. Concerning the FPWFS, it does converge and is as fast as the PWFS. These preliminary results are very encouraging with regards to the robustness of the FPWFS, a property that is extremely important in adaptive optics. \\

%To do so, we create turbulent phase screens following the Kolmogorov power spectral density with a $D/r_0=20$ moving at a speed of $10m.s^{-1}$. 
%The graph of the OPD residuals versus the number of loop iterations allows us to know if each WFS manages to boostrap and, if it does, the speed of this convergence. Simulations done on a large random draw of turbulence screens show that MPWFS converges and is the fastest WFS whereas the ZWFS does not manage to close the loop.
%Despite of its low linearity range, the PWFS turns out to boostrap effectively. Concerning the FPWFS, it does converge and is as fast as the PWFS. 
%This results prove the robustness of the FPWFS, property which is extremely important in adaptive optics. \\

This third performance paragraph will asses the behavior of the different WFSs in terms of \textbf{photon management}, i.e. the way each WFS is managing and optimizing each incoming photon with respect to each detector pixel. Note that this issue is central for low photon flux regimes which is usually the case for astronomical applications. 

A first approach to evaluate the WFS efficiency in dealing with low flux regimes is simply to evaluate how many photons are lost by the diffraction and the windowing. Table \ref{renta} shows these quantities. The PWFS only uses about 45\% of the incoming photons. This is due to the diffraction which rejects light outside the 4 pupils images, especially in the cross area between these images (see Fig. \ref{rappr}). The modulation corrects this by moving back the photons into the 4 pupil images with an efficiency of approximately 90\%. This value slowly increases with the amplitude modulation. The ZWFS performance is similar: still a 90\% efficiency, the photon being lost  due to diffraction effects.
In comparison, the FPWFS makes a better use of the incoming photons, with $\sim$98\% of them effectively used. This is the consequence of having removed the pupil separation.

\begin{table}[htb]
\centering
  \begin{tabular}{c|cccc}
   WFS & P & MP$_{3\lambda/D}$ & FP & Z \\
   \hline
Used photons (\%)& 43 & 88 & 98 & 85 \\
  \end{tabular}
\caption{Efficiency of photons.\label{renta}}
\end{table}

Secondly, the optical recombination of the 4 pupil signals operated in the FPWFS, requiring a smallest area on the detector compared to the PWFS, tends to prove that the FPWFS is particulary adapted to low photon flux regime. Indeed, for a same pupil sampling, 4 pixels are needed for the PWFS whereas only 1.1 is required for the FPWFS (for a 90\% overlap rate). As a consequence, the averaged flux per pixel will be about four times higher in the case of the FPWFS. This may be very interesting if the detector presents signficant electronic noise as it is the case for the near IR detectors.
Furthermore, for a WFS working at a visible wavelength -- where today detectors have sub-electronic noise -- the fact that the FPWFS needs less pixels than the PWFS may still be interesting. Indeed, in the context of Extreme AO (XAO), where a very large number of aberrations modes have to be mesured, a smaller detector can be very welcome since costs and data processing speed will be improved.

\medskip

A new WFS has been derived using the same optical design as the Pyramid WFS. The only modification is to consider a smaller Pyramid angle in order to let the pupil signals overlap. It allows an optical recombination of the four pupil images directly on the detector. %The resulting intensity turns out to efficiently code the phase information.

The new Flattened Pyramid WFS presents extremely promising performance. It appears in particular that this sensor conciliates two pillars of WFS: the sensitivity and the linearity range. More specifically, we demonstrated on simulations that: 
\begin{itemize}
\item At low radial orders (typically $<$4), the FPWFS provides the same noise propagation performance as a PWFS modulated with an amplitude of about 1.5 $\lambda/D$. At high radial orders, its behavior is strongly improved with almost a gain of one order of magnitude with respect to the PWFS and almost reaches the optimal ZWFS level. 
\item The linearity range of the FPWFS outperforms compared to other static WFSs as the PWFS or the ZWFS. And even if the FPWFS is not as effective as a large modulated MPWFS for the first Zernike radial orders, it becomes more linear for higher modes. It means that the FPWFS may not require any modulation.
\item We tested how the FPWFS could behave regarding real turbulence screens. As for the pyramid, and even if the linearity range is not as good as for other sensors, the FPWFS can effectively and efficiently close an AO loop.
\item Diffraction effects are reduced and photons which were lost between the pupil images for PWFS (and to a lesser extent for MPWFS) are now used to code phase information.
\item The illuminated area on the detector is smaller for the FPWFS than for the PWFSs. As a consequence, the new sensor requires less pixels to mesure a same number of aberrations, what makes of it a WFS particulary appropriate to high-contrast applications.

\end{itemize}

\noindent Acknowledgments: This work was co-funded by the European Commission under FP7
Grant Agreement No. 312430 Optical Infrared Coordination Network for Astronomy, by the ANR project WASABI and by the French Aerospace Lab (ONERA) (in the framework of the NAIADE Research Project).


\begin{thebibliography}{1}
\newcommand{\enquote}[1]{``#1''}

\bibitem{Sh71}
R.~B. Shack and B.~C. Platt, \enquote{Production and use of a lenticular
  hartmann screen,} J.\ Opt.\ Soc.\ Am. \textbf{61}, 656 (1971).

\bibitem{Wil75}
R.~G. {Wilson}, \enquote{{Wavefront-error evaluation by mathematical analysis
  of experimental Foucault-test data},} \ao \textbf{14}, 2286--2297 (1975).

\bibitem{Rag96}
R.~{Ragazzoni}, \enquote{{Pupil plane wavefront sensing with an oscillating
  prism},} Journal of Modern Optics \textbf{43}, 289--293 (1996).

\bibitem{Ndia13}
M.~{N'Diaye}, K.~{Dohlen}, T.~{Fusco}, and B.~{Paul}, \enquote{{Calibration of
  quasi-static aberrations in exoplanet direct-imaging instruments with a
  Zernike phase-mask sensor},} Astron. Astrophys. \textbf{555}, A94 (2013).

 
 \bibitem{Guy05}
O.~{Guyon}, \enquote{{Limits of Adaptive Optics for High-Contrast Imaging},}
  \apj \textbf{629}, 592--614 (2005).
 
\bibitem{Ver04}
C.~{V{\'e}rinaud}, \enquote{{On the nature of the measurements provided by a
  pyramid wave-front sensor},} Optics Communications \textbf{233}, 27--38
  (2004).

\bibitem{Rigaut92}
F.~{Rigaut} and E.~{Gendron}, \enquote{{Laser guide star in adaptive optics.
  The tilt determination problem},} Astron. Astrophys. \textbf{261}, 677--684
  (1992).
 

\bibitem{Bur06}
A.~{Burvall}, E.~{Daly}, S.~R. {Chamot}, and C.~{Dainty}, \enquote{{Linearity
  of the pyramid wavefront sensor},} Optics Express \textbf{14}, 11925--11934
  (2006).

\end{thebibliography}
\end{document}